\documentclass[12pt]{iopart}
\usepackage{graphicx}

\newcommand*\chem[1]{\ensuremath{\mathrm{#1}}}
\begin{document}

\title{The effect of boundary slippage  and nonlinear rheological response on flow of  nanoconfined water }
\author{ Amandeep Sekhon, V. J. Ajith, Shivprasad Patil }

\address{Nanomechanics Laboratory, Physics Division and Centre for Energy Science, h-cross,  Indian Institute of Science Education and Research, Pune 411008, Maharashtra, India
}
\ead{s.patil@iiserpune.ac.in}
\vspace{10pt}
\begin{indented}
\item[] 
\end{indented}

\begin{abstract}
The flow of water confined to nanometer-sized pores is central to a wide range of subjects from biology to nanofluidic devices. Despite its importance, a clear picture about nanoscale fluid dynamics is yet to emerge. Here we measured dissipation in less than 25 nm thick water films and it was found to decrease for both wetting and non-wetting confining surfaces. The fitting of Carreau-Yasuda model of shear thinning to our measurements implies that flow is non-Newtonian and  for wetting surfaces the no-slip boundary condition is largely valid.  On the contrary, for non-wetting surfaces boundary slippage occurs with slip lengths of the order of 10 nm.  The findings suggest that both, the  wettability of the confining surfaces and nonlinear rheological response of water molecules under nano-confinement play a dominant role in  transport properties. 
\end{abstract}
\section{Introduction}

Water permeation through hydrophobic channels, such as nanotubes, is five orders larger than expected from conventional fluid theory\cite{Mainak}. Similarly, Hydrophobic interiors of membrane proteins allow a rapid transit of water molecules\cite{Hille}. The flow of water through hydrophilic channels, on the contrary, is hindered compared to bulk water \cite{Derjaguin,Tas, Chuarev}. The measured viscosity through hydrophilic nanochannels is 30 percent larger than bulk water\cite{Tas}. The viscosity measurement of nanoconfined water by independent means have resulted in contradictory findings\cite{Zhu, Reido1, Raviv1, Khan1}. It has also been argued that water under confinement is a non-Newtonian fluid with finite relaxation\cite{Reido2, Jeffery, Khan, Karan2, Kageshima}. Despite contradictory claims, understanding flow behavior of water under nanoconfinement is important due to its relevance in a wide range of topics such as flow through biological pores\cite{Beckstein, Sansom, Granick1, Finney}, lubrication processes in nanomechanical devices\cite{Scherge}, filtration using nanoporous media\cite{Cohen, Lee} and transport through nanofluidic devices\cite{Tas,Chuarev}.

Conclusions regarding viscosity of nanoconfined water are difficult to reach owing to possible surface effects such as surface registry and finite slip at the boundary \cite{Zhu,Reido3}.  Although no-slip boundary condition is largely valid for bulk flows, it is suggested that there is a possibility of finite slippage when liquid flows through, or is squeezed out of the gap which is of the order of few nm\cite{Neto}. Indeed, to explain the enhanced flow rates in nanotubes, the violation of no-slip is invoked\cite{Mainak}. Researchers in the past have claimed contradictory findings regarding existence of slip for water nanoconfined by wetting surfaces\cite {Bonaccurso, Ducker}. The violation of no-slip boundary condition should reveal itself through reduction in measured stress. This affects the apparent viscosity of nanoconfined liquids and a parameter called ``slip-length'' is included in models describing the flow\cite{Reido3}. The slip-length, as defined by the Navier, is the distance from the boundary inside the solid where liquid velocity is extrapolated to be zero. If slip exists, a wetting surface is expected to exhibit smaller slippage compared to non-wetting ones.

Here, we measure the dissipation in nanoconfined water using a tuning fork based instrument developed in our laboratory.   A tip is oscillated over a surface at a distance of few nm and the intervening gap between the tip and surface is filled with water.  The measurement of change in dissipation from bulk to nanoconfinement is  measured when the tip approaches from bulk to a  distance of  $\leq 25$  nm from the surface. Eventually,  the tip makes contact with the surface draining all the liquid beneath it to the surrounding bulk. Figure 1 shows a schematic of the measurement. It is observed that the dissipation is reduced under confinement. Assuming that the change in viscosity under confinement is causing the reduction in dissipation, it  is fitted with  Carreau-Yasuda model of shear thinning along with finite slippage.    The characteristic time of shear thinning is found  to be $\sim 40  \mu$s.   To separate the effects of surface wettability  and  inherent slow-down in dynamics of water molecules on the flow of nanoconfined water, we performed measurements on substrates of different wettability.  We found that, for fully wetting to non-wetting substrates, Carreau-Yashuda model comprising finite slip-length fits better to experimentally observed reduction in dissipation. From the fitting we extract characteristic time-scales and the slip-lengths for all the surfaces. The slip is found to increase with  non-wettability ( larger contact angles), but characteristic time-scales of shear thinning do not vary significantly on different substrates.  Thus, the fitting of modified Carreau- Yashuda model to the data suggests that  it is possible to single out the dynamics of water molecules under confinement. The slip exists and it is negligibly small ($\sim$ 1.5nm)  for fully wetting surfaces ($\theta$ $\approx$ 0$^{\circ}$), such as mica,  and increases  upto $\sim$  12 nm  for intermediate ({$\theta$ $\approx$ 50$^{\circ}$}) and non-wetting surfaces ({$\theta$ $\approx$ 90$^{\circ}$}).  Although preliminary, these measurements could provide partial explanation of the enhanced permeation of water  through hydrophobic channels\cite{Mainak, Hille, Lee}.   The fitting exercise suggests that the flow of water at the nanoscale is determined by both  the slip-boundary and the altered response  from  Newtonian in bulk,   to rheological  under confinement.  The measurements also suggest that the flow under confinement should be treated as  that of complex fluids and different experiments  may provide inconsistent values for viscosity under confinement, largely determined by the operational parameters of the measurement.  

\section{Method} 
\subsection{The instrument}

 \begin{figure}

\includegraphics[width=1.0\linewidth]{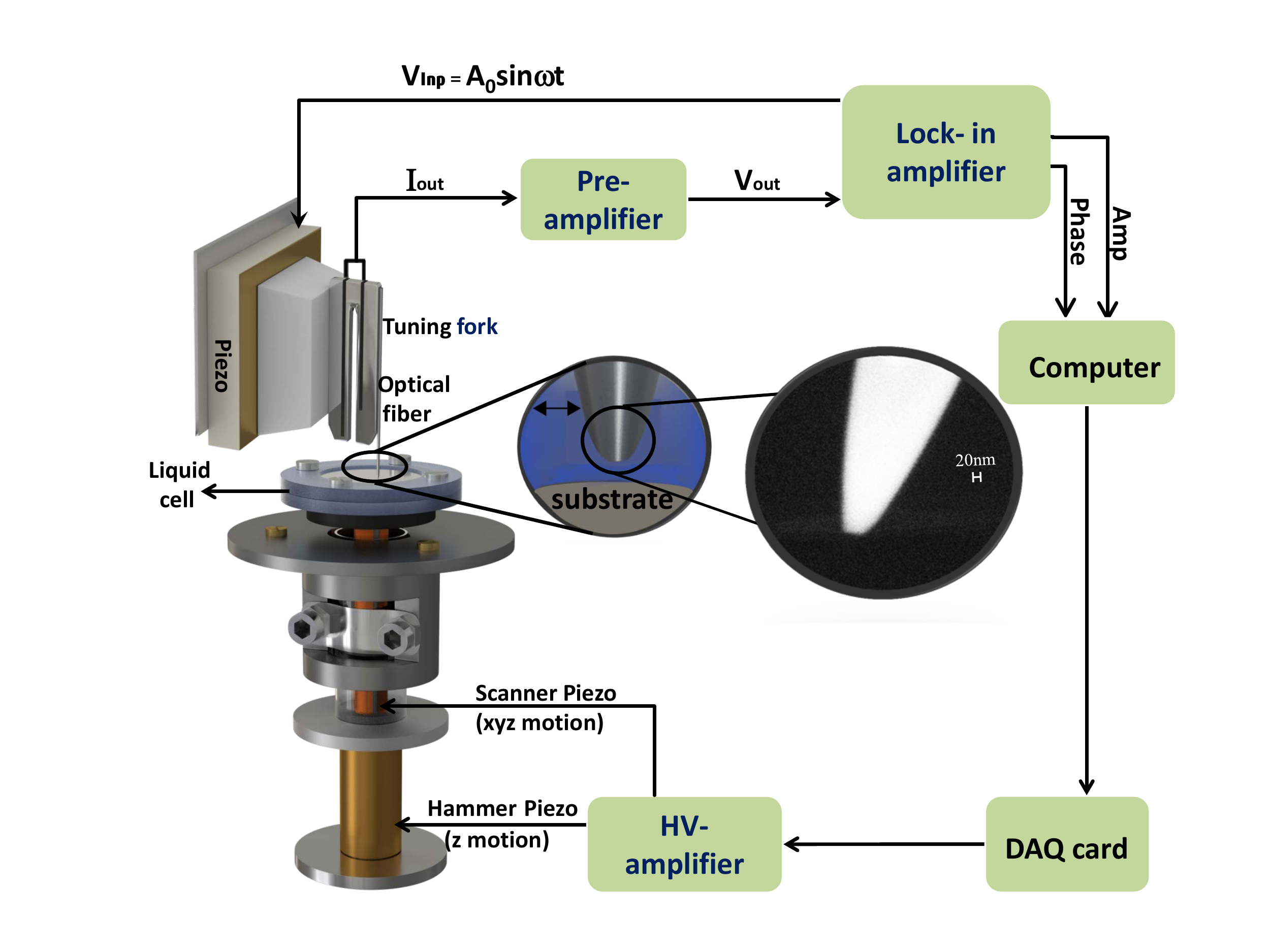} 
\caption{  Schematic of the measurements. A tuning fork is used as force sensor to measure the dissipation as the tip is oscillated off-resonance in liquid. The current through the electrodes connected to prong surfaces is a measure of the amplitude of the tip-bearing prong.  The tip is approached towards the substrate to cause the confinement. The change in dissipation due to altered flow response of the confined molecules beneath the tip is measured as a function of separation ($\leq$ 25 nm).      } 
\end{figure} 

The measurements  are performed using a home-built instrument. The schematic is shown in figure 1. A liquid cell is mounted on inertial-sliding nano-actuation stage. The bottom of the liquid cell holds the substrate which is one of the confining surfaces. The sample is approached towards the tip mounted on one of the prongs of tuning fork. 
Water is confined between the smooth surface and the fiber tip. The samples are kept in millpore water for few hours before placing them at the bottom of a liquid cell, which is again immediately filled with millipore water. In a typical measurement, the tip is oscillated off-resonance and the surface brought close to it in a controlled manner. The tips are prepared by pulling an optical fiber in laser-based fiber puller and are imaged under Scanning Electron Microscope (SEM) before use. Figure 1 shows schematic of water confinement in these experiments. The fiber-tip is fixed on one prong of the tuning fork and the other prong is mounted on a piezo which is used to provide off-resonance drive to the tuning fork. The current through tuning fork electrodes is due to differential bending in prongs. In off-resonance operation, amplitudes of both the prongs are equal and the prong motions are in phase. Therefore, the current through the electrodes due to bending is zero. The amplitude of  one prong which is  bearing the tip reduces due to viscous drag once the tip is immersed in liquid.   This causes a finite amount of current to flow through the electrodes. The change in current is measured as the tip is approached towards the surface using a lock-in amplifier and is used to estimate the change in amplitude. This, in turn, is a measure of dissipation in the tip oscillations.  Over the distance covered by the tip in our experiments, the bulk drag force changes by immeasurably small amount. The changes in dissipation are  measurably large once the tip is within few nm from the surface with water in the intervening gap.

\subsection{Measurement of shear amplitude} 
The instrument is used to measure the change in dissipation of tip oscillations when water is confined beneath it.   The instrument works in off-resonance mode where the tuning fork bearing the tip is mechanically driven by piezo-drive.  In this section we describe the method of estimating the amplitude by measuring the current through the electrodes. 

For a mechanically driven tuning fork, the current through its electrodes is given by $ I = \alpha\omega ( A_{drive}- A_{tip} ) + I_p$. Where  $( A_{drive}- A_{tip})$ is the differential amplitude. $ A_{drive}$ - drive amplitude, $ A_{tip}$ - amplitude of the tip-bearing prong.  $\alpha$ is piezoelectric coupling constant for the tuning fork.  $I_p$ is current through parasitic capacitance.  When the tuning fork is driven off-resonance and the tip is in air, $A_{drive} = A_{tip}$ and the current due to differential bending is zero.  $A_{drive}$ is measured independently using fibre-based interferometer. The  bulk amplitude $A_b$,  when the tip is immersed in liquid and  is  close to the substrate ($\sim 25$ nm),  is determined as follows.  Since the differential bending of the prongs is zero  in off-resonance conditions,  the current when tip is in air $I_a$ is purely the parasitic current $I_p$; $I_a = I_p$.   Similarly, $I_b = \alpha\omega ( A_{drive}- A_b ) + I_p$; where $I_b$ is the current when tip is about $25$ nm from the substrate. The difference in the current $\Delta I$ gives the amplitude $A_b$ through the following equation

\begin{equation}
\Delta I = I_b-I_a = \alpha\omega ( A_{drive}- A_b) + I_p - I_p = \alpha\omega (A_{drive}-A_b) \nonumber 
\end{equation}

 The value of $\alpha$ is $ 12 $ $\mu$C/m. The amplitude when the tip is close to the surface ($\sim$ 0 - 25 nm ) and  water is confined with altered flow properties is determined in a similar manner.  If $I_b$ is current corresponding to amplitude $A_b$ and $I_c$ is current corresponding to the tip amplitude $A(d)$ when liquid is confined beneath it, then

\begin{equation} 
\Delta I_c = I_c - I_b = \alpha\omega (A_b - A(d))\nonumber 
\end{equation} 
The measurement of $\Delta I_c$ gives the amplitude $A$ of the prong when the water is confined beneath the tip. The measurement $\Delta I$ gives  the amplitude of the tip in bulk ($A_b$) through equation 1.    

Note that, in both cases the positive $\Delta I$ corresponds to increase in the tip dissipation.  The observation in our experiments is that when the tip is immersed in liquid from air,  the $\Delta I $ is positive.  This is due to the enhanced viscous drag  experienced by the tip being in contact with liquid.
 When the tip is close to the substrate, the $\Delta I_c$ is negative. This indicates a reduced tip dissipation under confinement  compared to bulk.

\subsection{Dissipation from shear amplitude} 
The dissipation is characterised by the damping constant $\gamma$.  The changes in amplitude and phase  of an oscillating tip in liquid can be used to estimate the  damping constant.  This dissipation has two contributions, a) the hydrodynamic drag force acting on the tip due to surrounding liquid and b) the altered flow properties of liquid under confinement. The overall hydrodynamic drag on the tip does not change during our experiments (The tip area and the liquid used in the experiment is same). The change in amplitude is purely due to the changes in liquids response under confinement. In the following, we discuss the method used to  measure change in dissipation under confinement. 
In off-resonance conditions, 
\begin{equation} 
\gamma = -k{A_0\over{A}\omega}sin\phi
\end{equation}

Where $ A_0$ is free amplitude without the dissipation which is equal to drive amplitude.  $\phi$ is  
phase lag between the drive and the tip-end  in presence of dissipating medium.

If $A_b$ is the amplitude when the tip-sample separation is about 25 nm, $A_d$ is drive amplitude  and $\phi_1$ is the phase lag due to immersion of tip in the liquid from  the air. The dissipation in bulk $(\gamma_b)$ which is largely due to hydrodynamic force is 

\begin{equation} 
\gamma_b = -k{A_d\over{A_b}\omega}sin\phi_1
\end{equation}
Where $k$ is the stiffness of the prong. The dissipation,  when the tip is close to the substrate ($\leq 25$ nm)  and water is in confined state characterised by the separation dependant amplitude $A (d)$  is given by
\begin{equation}
\gamma_c(d)   = -k {A_d\over{A(d)}\omega}sin(\phi_1 + \phi_2)
\end{equation}
Where $\phi_2$ is the additional phase lag from bulk to the confined state. This was found to be zero in our experiments. 
The relative change  in dissipation using equations 4 and 5 
\begin{equation} 
 { \Delta \gamma} / \gamma_b  =    \gamma_c  / \gamma_b - 1  =    A_b/A(d) - 1 
 \end{equation}
 
  $A(d)$  is measured as the tip is approached towards the surface to obtain the change in dissipation compared to the bulk dissipation. 
  
  To validate our methodology of estimating dissipation from the amplitude measurement, we measure dissipation for two organic liquids of known viscosity. One is octamethylcyclotetrasiloxane (OMCTS), a model lubricant having viscosity 2.6 cP, the other liquid is heavy liquid paraffin  with considerably high viscosity (55 cP) compared to OMCTS. We took the same volume of liquid in the liquid cell. This ensures that the level of liquid above the bottom surface is the same for both liquids. The tip is then immersed in it and approached till it reaches the surface through auto-actuation without breaking the tip. The tip is then pulled back by 100 nm. The process ensures that the same length of the fiber tip ($\approx$ 0.75 mm) is immersed in the liquid for both OMCTS and paraffin. Under these circumstances the area of the tip pushing against the liquid while it is oscillated in it is the same. As mentioned earlier, the ratio of dissipation in OMCTS and paraffin should be equal to the ratio of their viscosities. We measured the amplitude ($A_b$)and the phase lag ($\phi_1$) at different drive amplitudes and frequencies which are far below resonance for both liquids and calculated the dissipation using equation 4 for each of them. This ratio turned out be $ 20.68 \pm3$. The ratio of viscosities of paraffin and OMCTS is $21.2$. This ensures that we are measuring dissipation constants accurately with our method. 
 
 \subsection{Models}

 The observed reduction in dissipation close to the substrate($ \leq$ 25 nm), can be attributed to  a) slip-boundary, b) shear thinning  and c) combined effect  of slippage and shear thinning.  In the following we discuss the models to describe each of the three different scenarios. 
 \subsubsection {Slip boundary:}

  No-slip boundary condition is likely to be violated when one is measuring the flow properties in nanochannels\cite{Neto, Bonaccurso}. This slippage is characterised by a slip length $L_s$, which is a distance from the boundary inside the solid where liquid velocity is extrapolated to be zero. 
 At separation $d$,  the shear rate for no-slip boundary condition is $A\omega/d$  and  for finite slippage it  becomes  $A\omega/(d+L_s)$.   The viscous stress for newtonian liquids is given by the viscosity multiplied by the shear rate. If the liquid under confinement retains its Newtonian nature of rate independent viscosity, the finite slip reduces the viscous stress  due to altered shear rate. The ratio of dissipative forces  for slip and no-slip boundary is then given by the ratio of shear rates.  Since the force is proportional to damping constant,  the altered dissipation coefficient of nanoconfined water due to finite slippage is given by  
  \begin{equation}
\Delta \gamma/ \gamma_b = \gamma_c/\gamma_b -1 =  d/(d+L_s) - 1 
\end{equation}   
Where $\gamma_c$ is dissipation coefficient under confinement ($\leq$ 25 nm) and  $\gamma_b$ is dissipation coefficient in bulk. 
\subsubsection{Shear thinning:}

The other possible reason behind reduction in dissipation upon confinement is shear thinning. Here the change in dissipative force experienced by the tip can be related to altered viscosity. In shear thinning liquids,  this depends on shear rate. The phenomenon is observed in non-Newtonian fluids such as whipped cream, polymer melts and colloidal suspensions. It can be described using Carraeu-Yasuda Model which relates reduction in viscosity to shear rate\cite{Carraeu}. \begin{equation}
\eta = \eta_0[1+(\dot\Gamma{\cdot}\tau)^{a}]^{(n-1)/a}
\end{equation}
Where, $\dot\Gamma$ is shear rate, $\tau$ is characteristic shear thinning time and $n$ is exponent of the power law region of shear thinning and should be between 0 and 1, $a$ is parameter that characterises the transition from the Newtonian region to the power law. The shear rate in our experiments is given by $\dot\Gamma = A_0\omega/d + rv/2d^2 $. The term $rv/2d^2$ is due to the rate of water squeeze out  and depends on the speed $v$  with which we bring the tip towards the surface \cite{Khan1}.  In our experiments the typical values are $r = 25$ nm, $A_0 = 3 $ nm and $\omega =  2\pi \times 15\times 10^{3} $ rad./s. For $d= 0.1 - 10$ nm   The first term $A_0\omega/d $ becomes ~ $10^{4}$ to $10^{6} s^{-1}$.   For highest approach rates $v =  3 $ nm/s,  the second term $rv/2d^2$ becomes $10$ to $10^{3}$ $s^{-1}$. This means that contribution from the squeeze-out to shear rates is three orders smaller than the oscillatory shear. We can ignore the second term in further analysis. The fit procedures also show that the inclusion of the second term in the shear rate does not alter the fits significantly.   Equation 8 becomes

\begin{equation} 
\Delta \eta / \eta_0 ={\eta - \eta_0\over\eta_0} = [{1 + (\tau A_0\cdot\omega/d) }^{a}]^{(n-1)/a } -1 
\end{equation}
For $L_s = 0$, this relative change in viscosity $\Delta  \eta/ \eta_0$  is equal to relative change in dissipation coefficient $\Delta \gamma/ \gamma_b$


\subsubsection{Shear thinning with slip boundary:}  The third possibility is that the reduced dissipation close to the substrate is a combined effect of both slip-boundary ($L_s\neq 0$) and shear thinning.   We replace $d$ by $d+L_s$ in equation 9 to include slippage in Carreau-Yasuda model.

\begin{equation} 
\Delta \eta / \eta_0 ={\eta - \eta_0\over\eta_0} = [{1 + (\tau A_0\cdot\omega/(d+L_s) )}^{a}]^{(n-1)/a} -1 
\end{equation}

The relative change in dissipation measured experimentally can be related to the viscosity since the tip-geometry remains the same, $\Delta \eta / \eta_0 = \Delta \gamma /\gamma_b $.  We fit the expression on right hand side of 7, 9 and 10 with the measured reduction in dissipation given by the right hand side of equation 6 
\subsection{Sample preparation} 
  Water is confined between the substrates of different wettability and a fiber tip. The fiber tip having diameter $\sim 50$  nm is prepared by pulling the single mode optical fiber in  fiber-puller having a  CO2 laser (Sutter Instrument Co. P2000).The wettability  of the samples is determined through contact angle measurements. We used five different substrates, mica ($\theta$  =  5$^{\circ}$), Silicon Carbide SiC ($\theta$  =  42$^{\circ}$), Aluminium Oxide  \chem{Al_2O_3} ($\theta$  =55 $^{\circ}$), Lanthanum Oxide LaO ($\theta$=65$^{\circ}$) and Hydrogen terminated silicon ($\theta$=75 $^{\circ}$). Mica  is  freshly cleaved with a scotch tape  and then placed in the liquid cell. The liquid cell was immediately filled with MilliQ water.  Single crystals of SiC, \chem{Al_2O_3} and LaO  were  first rinsed with ethanol  followed by sonication in ethanol    and MilliQ water for 10 minutes  each.  Hydrogen terminated Silicon substrate is prepared by dipping single crystal Si substrate in Hydrogen Fluoride (HF) solution for 5-10 minutes  to terminate surface with hydrogen and rinsed with water.  Each substrate is kept in  MilliQ water for few hours to equilibrate before the experiments.

 \section{Results} 
 \begin{figure}
 \includegraphics[width=1.0\linewidth]{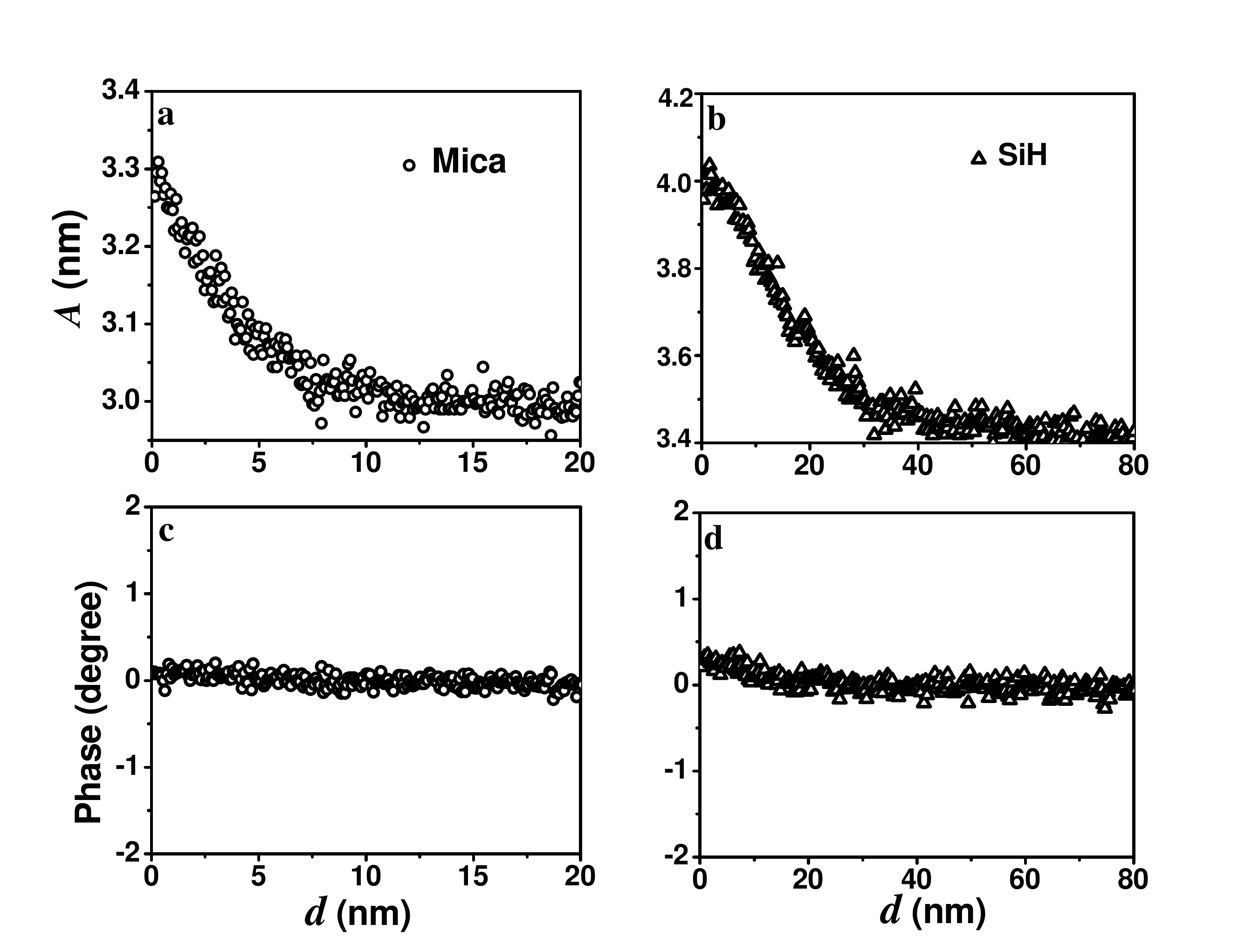} 
 \caption{ The measured amplitude of the prong bearing the tip in off-resonance measurement.  The amplitude starts to deviate from bulk for separations below 10 nm for mica(a) and 25 nm for Si-H (b). On both substrates the amplitude of the prong increases compared to bulk indicating reduction in dissipation. The amplitudes are calculated from measured current using equations 1 and 2.   The  corresponding phase measurements are shown in (c) and (d). The phase does not vary under confinement allowing the use of equation 6 to compute dissipation.       } 
 \end{figure}
 
 \begin{figure}

\includegraphics[width=1.0\linewidth]{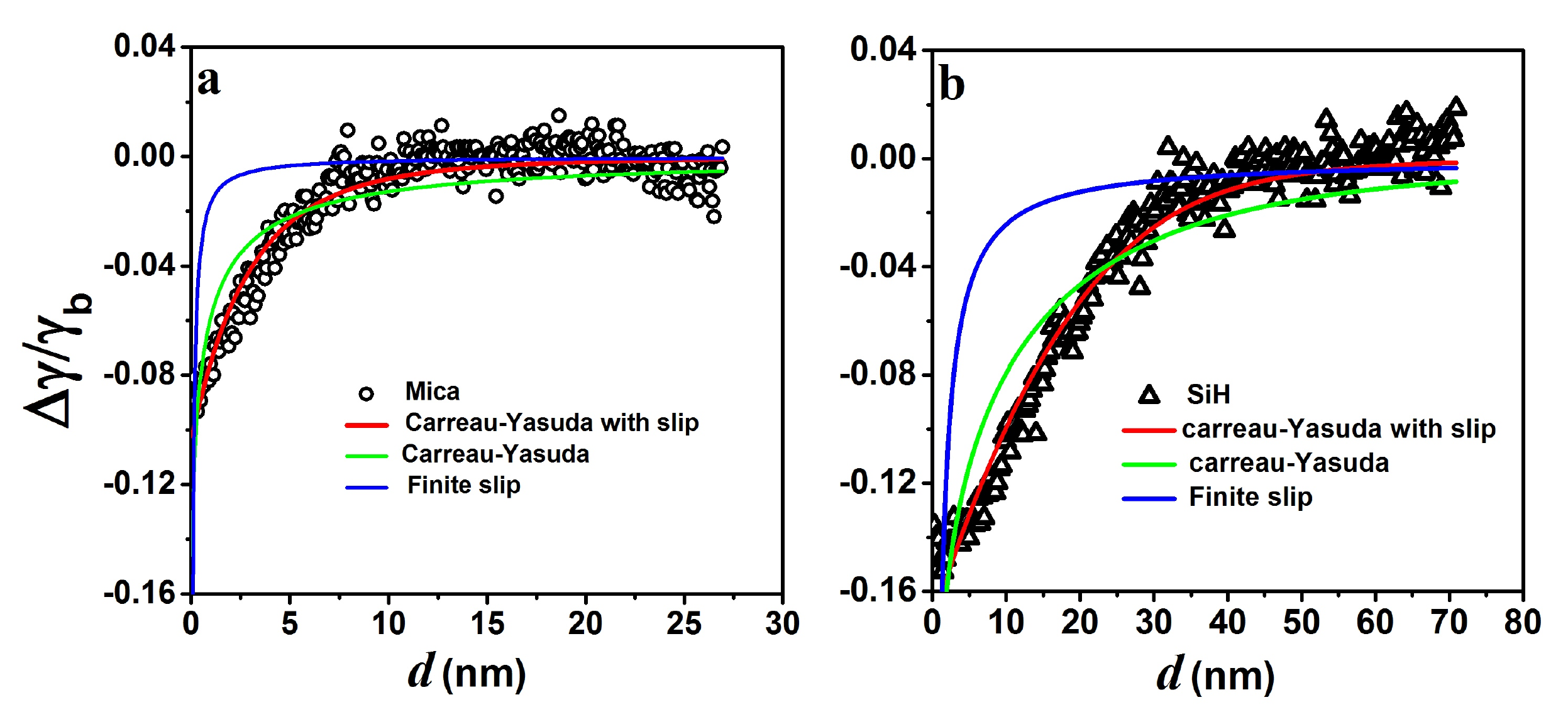} 
\caption{ The measurement of change in dissipation as the tip approaches the surface of  (a) mica  and (b) hydrogen terminated Si surface. The open circles and triangles are experimental data  and continuous lines are fits of various models.  Blue line - finite slip and no change in water viscosity,  Green line -  viscosity changes as per the Carreau-Yasuda rule without slippage and   red line - the modified  Carreau-Yasuda with finite slippage. Clearly, the modified Carreau Yasuda fits better than other two models for both  wetting ( mica - $\theta$ $\approx$ 5$^{\circ}$ ) and non-wetting surfaces ( SiH - $\theta$ $\approx$ 75$^{\circ}$ ). Table 1 shows all the fit parameters of fitting equation 10 to experimental data.  The fitting exercise shows that the dynamics under confinement is characterised by both nonlinear rheological response akin to complex fluids and finite slippage at boundary.   The relevant fit parameters for different wettability are plotted in figure 4.} 
\end{figure} 

 Figure 2 (a) and (b)  shows the measured amplitude  of the tip-bearing prong versus separation on mica and Si-H respectively. Figure 2 (c) and (d) are  corresponding phase values. Since the phase lags are zero in all the measurements, equation 6 can be used to estimate dissipation where $\phi_2$ is taken to be zero. The zero of the separation in AFM methods is determined by a point at which the normal deflection of the lever suddenly changes\cite{Khan}. It is assumed that the last layer  remains bound and has the same shear response under varying loads\cite{Zhu}.  In our case, the zero is determined by the point at which the shear amplitude does not vary any further. For a wetting substrate (mica) the amplitude starts to increase below 10 nm  and for non-wetting substrate(SiH) below 25 nm.  Qualitatively,  the data indicates reduced  dissipation under confinement.  The data shown here is representative.  Typically more than  20  measurements are performed on each substrate in a range of frequency ($10 -15$  KHz) and amplitude ($1 - 3$ nm).  Figure 3 (a) and (b) show relative change in dissipation calculated  from data in  figure 2(a) and (b) using equation 6. The dissipation under confinement is reduced compared to the  dissipation in bulk. This reduction is $\sim$ 10  \%.  Note that  this 10 \% change is compared to the dissipation due to the entire tip. The entire tip area is roughly  $\sim 10^{-7}  m^{2} $    ;  the area at the end of the tip which serves as one of the confining surfaces is $\sim 10^{-15} m^2$.    The change in dissipation is entirely due to  an altered flow response of water molecules beneath the tip which is referred to as  nano-confined.   This change is 10 \% of the total dissipation due to  macroscopic tip moving in the bulk water.  This shows that  the  altered flow response due to confinement  is quite significant.   
 
   In the following we explore the possible reasons behind reduction in dissipation when water is confined beneath it.  If no-slip boundary is violated retaining the viscosity to bulk value,  the slippage can result in reduced viscous drag. The other possibility is shear thinning. Here,  the change can be related to altered viscosity which depends on shear rate.  The observed  dissipation reduction can also be attributed to the combined effect of shear thinning and finite slippage. All three possible scenarios are discussed in details  in the section "Methods".    
    
   Figure 3 (a) and (b) show fits of models given by equations 7, 9 and 10  to experimental data on mica and Si-H respectively.  The green continuous line is fit of  Carreau-Yasuda model of shear thinning to the data given by equation 9. The blue continuous line represents finite slippage described by equation 7. The red line is modified Carreau-Yasuda to include finite slip. The comparison between three fits clearly shows that the observed dissipation can be attributed to both the nonlinear rheological response of shear thinning at high shear rates ($10^{6} s^{-1}$)   and finite slippage at the boundary.  To elucidate the role of surface wettability on the dynamics of water confined at nanoscale, we repeated the measurements on five substrates with different degree of wettability characterised by the contact angle (data not shown). We used mica ($\theta$  =  5$^{\circ}$), Silicon Carbide SiC ($\theta$  =  42$^{\circ}$), Aluminium Oxide  \chem{Al_2O_3} ($\theta$  =55 $^{\circ}$), Lanthanum Oxide LaO ($\theta$=65$^{\circ}$) and Hydrogen terminated silicon ($\theta$=75 $^{\circ}$).   In all these measurements it is seen that the modified Carreau-Yasuda with finite slippage fits better than  pure shear thinning or pure slippage. 
  
 Figure 4(a) and (b)  show a plot of characteristic slip length $L_s$ and shear thinning time-scale $\tau$ for different substrates. It is plotted versus the contact angle. The errors are estimated from 20 different measurements on each substrate. To  examine the robustness of fits, we imposed  shared initial values of the parameters for all substrates. We found that the fits converge with fitting parameters reported here.   By fixing the values of $L_s$ and $\tau$ obtained for mica  to fit  Si-H  data, or values from Si-H to fit mica data,  we could not get good fits to experimental data in either case.  These parameters were then freed and we see that they converge with values reported in the table.  For all the fits we kept a bound of 0-1 for the exponent $n$. The  $\tau$ varies from 28 to 64  $\mu$ s  for wetting to non-wetting surfaces,  a two fold increase.   The slip length for substrates with different wettability follows a expected trend. The slip  progressively increases from,  less than 1.5 nm  for wetting (mica),   to 12 nm for  non-wetting substrates (Si-H),  a  six fold increase. Considering that the errors on the relaxation are significantly larger  compared to the slip-length, we can claim that  the fitting of  modified Carreau-Yasuda model to the experimental data successfully separates the effect of surface wettability on the dynamics under confinement. Moreover, The characteristic shear thinning time-scale matches well with other similar observations wherein, the dynamics of water condensed over a wetting substrate is probed with on-resonance  method\cite{Jhe2}.
   
 \begin{figure}

\includegraphics[width=1.0\linewidth]{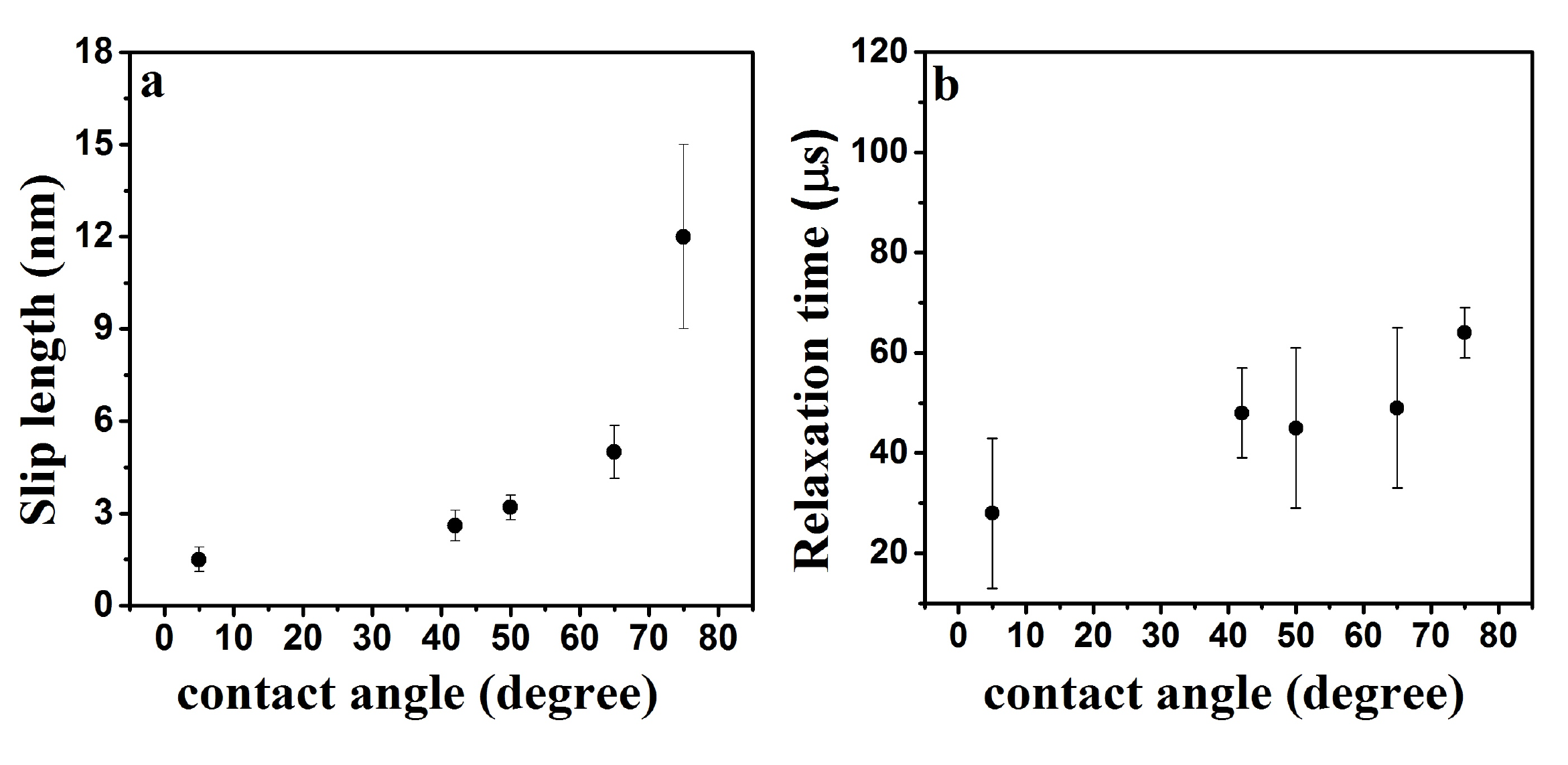} 
\caption{ (a) The slip length $L_s$ versus the contact angle for five different substrates. The slip length is below 1.5  nm for wetting substrate such as mica( $\theta$ =  5$^{\circ}$ ) and reaches upto 12 nm for SiH ( $\theta$ =  75$^{\circ}$ ). ( b)  The characteristic shear thinning time scale for all substrates. There is no appreciable change over  contact angles from  $\theta$ =  5$^{\circ}$ to 75$^{\circ}$. The error bars are estimated from 20 different measurements on each substrate.   } 
\end{figure} 
 
 \begin{table}[ht]
\caption{Fit parameters obtained on five substrates for Carreau-Yasuda model with finite slippage}
\centering
\begin{tabular} {c c c c c }
\hline \hline 
substrates\ & $\tau$($\mu$s)\ & $L_s$ (nm)\ & n\ & a \\ [1ex]
\hline
Mica & $28\pm15$ & $1.5\pm 0.4$ & $0.77\pm0.16$ & $2.3.\pm0.5$\\ 
SiC & $48\pm9$ & $2.6\pm0.5$ & $0.75\pm0.06$ & $2.9\pm0.7$ \\ 
Al$_2$O$_3$ & $45\pm16$ & $3.2\pm0.4$ & $0.76\pm 0.07$ & $2.5\pm0.5$ \\
LaO & $49\pm16$ & $5\pm0.8$ & $0.8\pm0.1$ & $2.9\pm1.1$ \\
Si-H & $64\pm5$ & $12\pm3$ & $0.89\pm0.02$ & $2.5\pm0.3$ \\ 
 [1ex] 
\hline
\end{tabular}
\end{table}

We report all the fitting parameters ( $\tau$,  $L_s$,  a and n) for different substrates in table 1.  It is noteworthy that,  the  other two parameters $a$ and $n$ related to Carreau-Yasuda model of shear thinning do not vary with wettability of substrate.   We remark here that Carreau-Yasuda with finite slip describes the  flow behaviour of the confined water accurately. Moreover, the method  proposed here of separating the effects such as,  finite slippage and slow-down in dynamics due to molecular  microstructure  is quite robust. The fit parameters related to Carreau-Yasuda does not change with surface wettability, whereas the parameter $L_s$ associated with finite slippage increases with it, a familiar relationship between wetting and slippage\cite{Neto}. The values for slip-length and its dependence on contact angle is  qualitatively consistent with quasi-universal relationship between the contact angle and slippage obtained from molecular dynamics simulation by Huang et al.\cite{Huang}     
 
 \section{Discussions}
 
Li et al. measured the viscosity of nanoconfined water to be several orders of magnitude higher than bulk\cite{Reido1}.   They have also reported nonlinear viscoelastity of confined water\cite{Reido2}.   On the contrary, the measurement by Raviv et al.  suggested that water viscosity under confinement remains close to bulk or reduces by three times the bulk viscosity\cite{Raviv1}. Our novel dynamic shear measurement method concludes that dissipation offered by the water confined under the tip is reduced. The change is 10 \% of the tip-dissipation.  This is a considerable change given the area of the macroscopic tip immersed in water compared to the confinement area. We attribute the dissipation reduction  under confinement to shear thinning with characteristic time-scale of $ 40 \mu$s and a finite slip (1 - 12 nm).  The characteristic shear thinning time-scale is roughly of the same order as Maxwell's relaxation previously measured\cite{Karan2}.  In order to compare our results  with shear measurements from groups of Reido\cite{Reido1} and Klein\cite{Raviv1}, we need to perform experiments at low shear rates and frequencies.  It is difficult  to extend our measurements to frequencies of the order of  1 KHz. This is because the current through the tuning fork prongs depends on frequency and amplitude. For frequencies below 1 KHz, the response is immeasurably low.   The shear measurements  by Riedo similarly will be difficult in higher frequency owing to poor quality factor ($\approx$  1) in liquid environments. The two methods are complementary in terms of parameter space in which they work.
 
 For high shear rates using torsional AFM measurements, a shear rate-dependant response is observed for less than 1 nm thick  water films\cite{Kageshima}  Recently, shear thinning is also reported using a  on-resonance tuning fork based shear apparatus by Bongsu et al.\cite{Jhe2} and the characteristic shear thinning time scale is reported to be 1  $\mu$s.  These results, together with our observations highlights  a need for a  development of a unified picture based  on operational parameters of different methods.  
 
 Why do we observe  reduced dissipation close to the substrate as opposed to enhanced viscosity reported  by many others?   It is well known that the  response of a rheological material varies with  strain parameters.    For  $\dot{\Gamma}  \tau$ much less than 1, the system response is Newtonian.  For relaxations closer to bulk value ($\approx$ 1 ps)   the experimentally accessible shear rates and frequencies are much smaller than the inverse of the relaxation time. The response is then bound to exhibit a Newtonian behaviour.  Shear thinning is observed for shear rates where $\dot{\Gamma} \tau  \textgreater 1 $. Indeed,  it is observed in experiments where the liquid under confinement is sheared faster than the relaxation\cite{Kageshima, Granick}. The amplitudes (1-3 nm) and frequency range (10-15 KHz) used in our measurement together with relaxation (40 $\mu$s) obtained from fits to Carreau-Yasuda  shows that,  $ \dot{\Gamma}  \tau=   0.5$ to $5$.   Our operational parameters  lie in the range where shear thinning is observed. It also suggests a need for independently determining relaxation of confined water.

The nanofluidic measurements on flow of water suggest that water viscosity increases slightly in hydrophilic channels\cite{Tas, Chuarev}. The flow through carbon nanotubes (hydrophobic channels), on the other hand is four to five orders faster than the one expected from fluid dynamic equations. The permeability through membranes is shown to enhance with decrease in pore size\cite{Lee}. The reasons behind such rapid flow are largely unknown and are usually attributed to enormous slippage at the boundary\cite{Mainak}. Here we show that not only slip is responsible for enhanced flow but shear thinning also plays a significant role. Our measurements imply that flow through nanochannels can be complex and may result in much larger flux with reduced viscosity under high shear rates. The flow rate will be determined by both, the phenomenon of shear thinning and the slippage at the boundary that depends on wettability of confining surfaces.

The existence of slip has been historically debated \cite{Neto}. In 1845, G. Stokes, based on the experiments at the time and his own calculations concluded that no-slip boundary condition is valid for all flows. In recent years, the attempts are being made to determine slip from fitting Reynold's equation to hydrodynamic drag experienced by a sphere in front of plane using AFM. This has resulted in contradictory findings about the existence of slip\cite{Neto, Bonaccurso, Ducker}. In small-amplitude AFM measurements of squeeze-out dynamics, conclusion regarding slip could not arrived at \cite{Khan1}. Ortiz-Young et al. have fitted oscillatory shear data using an AFM to a model based finite slip \cite{Reido3}. The measurements on wetting ($\theta$ $\approx$ 0$^{\circ}$), intermediate ({$\theta$ $\approx$ 50$^{\circ}$ }) and non-wetting ({$\theta$ $\approx$ 90$^{\circ}$ }) samples in our study conclude that although slip length is close to zero  for fully wetting surfaces it takes values of the order of 10 nm for non wetting surfaces. This may play a crucial role in determining the discharge of water through carbon nanotubes or hydrophilic nanochannels.

Shear thinning is commonplace in binary mixtures, polymer melts and in colloidal suspensions. Its existence in case of pure water confined to small dimensions is puzzling. It is usually attributed to forming and breaking of flow induced microstuctures\cite{Vermant}. Recently, shear thinning was shown to have entropic origins by directly imaging the suspension with fast confocal microscopy\cite{Cheng}. At the moment, it is difficult to point out origins of shear thinning of pure liquids under nanoconfinement. Further experiments planned in our laboratory to simultaneously measure stress and diffusion by means of optical spectroscopy could provide molecular level explanation of the phenomenon observed here. 

                Shear thinning has been observed for dodecane liquid in confinement\cite{Granick}. A clear and quantitative evidence of shear thinning in case of nanoconfined water reported here suggests a general resemblance in behaviour between organic solvents (non-polar, non-associative) and water (polar and associative). These were thought to behave differently under nanoconfinement in the past \cite{Raviv1}. Indeed, the flow through carbon nanotubes also shows faster flow rates for both water and organic solvents\cite{Mainak}. This hints at a need for a general understanding of reasons behind nonlinear rheological response of pure liquids under nanoconfinement.

 It has been argued  that reduction in water viscosity,  as opposed to the enhanced viscosity of organic solvents under confinement,  is due to the breaking of hydrogen bonds in water under confinement.  Our normal stiffness and damping measurements in the past have shown that the polar water and OMCTS behave in a similar manner under confinement and both show dynamic solidification \cite{Khan, Patil}.  The slow down in dynamics from ps to more than $\mu$s is  suggested to be arising out of criticality of nanconfined water related to a second order phase transition of capillary condensation with respect to pore size \cite{Karan2}. We emphasize   a need for simultaneous spectroscopic and stress measurement for divulging the molecular origins of slow relaxation and reduced dissipation in nanoconfined water.    
 \section{Conclusions}  
 
In summary, we have performed dynamic shear measurement on water confined between a sharp tip and substrates of different wettability. We explain the experimental observation of reduction in dissipation under confinement with the help of Carreau-Yasuda model of shear thinning and/or finite slippage at the boundary.
We found a clear evidence for shear thinning along with finite slippage for both wetting as well as non-wetting substrates. The slip length extracted from the fit procedures progressively increase for non-wetting substrates. On the contrary, the shear thinning time scale does not vary appreciably over five substrates with different degree of wettability.   The method allows the  separation of  contributions  arising out of  surface wettability and slow-down in molecular dynamics.  The findings have  relevance  in understanding the flow in nanofluidics and explaining rapid transit of water through carbon nanotubes reported earlier\cite{Mainak, Lee}.

\end{document}